\documentclass[prl,showpacs,twocolumn,amsmath,amssymb,floatfix,superscriptaddress]{revtex4-2}
\usepackage{graphicx} 
\usepackage{bm}
\usepackage{bbm}
\usepackage{color}
\usepackage{amsmath}
\usepackage{amssymb}
\usepackage{color}
\usepackage{footnote}
\usepackage{comment}
\usepackage{hyperref}
\usepackage{subfigure}
\usepackage{appendix}
\usepackage[applemac]{inputenc}

\begin{document}
\title{Two-Cooper-pair Josephson junction from Cooper-pair splitting}

\author{Pei-Hao Fu}
\email{peihao.fu@unifi.it}
\affiliation{Department of Physics and Astronomy, University of Florence, I-50019 Sesto Fiorentino, Italy}

\author{Luca Chirolli}
\email{luca.chirolli@unifi.it}
\affiliation{Department of Physics and Astronomy, University of Florence, I-50019 Sesto Fiorentino, Italy}

\date{\today} 
\begin{abstract}
Two-Cooper-pair Josephson junctions are key elements for parity-protected superconducting qubits and for Josephson circuits with tailored harmonics. 
We propose a device that realizes such a junction via Cooper-pair splitting. Four quantum dots between two superconducting leads sequentially split two Cooper pairs. A first pair of quantum dots operates as a Cooper-pair splitter, effectively suppressing local Andreev reflection, and a second pair of quantum dots operates as a buffer to host the first split pair, allowing a second pair to split. For infinite charging energy $U$ and superconducting gap $|\Delta|$ the current is purely $\sin(2\varphi)$, while single-pair corrections $\propto\sin(\varphi)$ scale as $1/U$ and $1/|\Delta|$. By threading a magnetic flux through the device and by tuning the asymmetry of the junction, we show that a Josephson diode effect arises, whose efficiency as a function of the flux makes it possible to identify the dominant harmonic of the current, thus providing a diagnostic tool accessible in dc measurements. 
 \end{abstract}
\maketitle

{\it Introduction.---}
Two-Cooper-pair junctions are a key ingredient for designing parity-protected superconducting qubits  \cite{ioffe1999,ioffe2002topologically,ioffe2002possible,doucot2003topological,doucot2005protected,kitaev2006protected,doucot2012physical} and engineering Josephson circuits characterized by multiple harmonic content \cite{chirolli2021enhanced,chirolli2022swap,gyenis2022,souto2022,calzona2023,leblanc2025,banszerus2024voltage-controlled,banszerus2025hybrid,chirolli2026fraxonium}. 
Such a junction shows the $h/4e$ Josephson effect, and its energy as a function of the superconducting phase difference $\varphi$ has the form $\propto\cos(2\varphi)$, featuring two degenerate but distinct minima within a $2\pi$ period. 

The established route to a two-Cooper-pair junction relies on a superconducting quantum interference device (SQUID) with two Josephson junctions per arm \cite{blatter2001design}, threaded by half a flux quantum $\Phi_0/2$, with $\Phi_0=h/2e$ the superconducting flux quantum. Destructive interference  suppresses single-Cooper-pair tunneling and leaves only processes involving pairs of Cooper pairs, in analogy with the interference-correlated pair-superfluid phase  \cite{doucot2002pairing,rizzi20064e-condensation,rizzi2006phase,cartwright2018rhombi-chain,chen2025} arising from Aharonov-Bhom caging    \cite{vidal1998ABcages}. Implementations based on this interference mechanism have been studied theoretically and realized experimentally in Josephson-junctions networks  \cite{pop2008measurement,gladchenko2009superconducting,bell2014protected,bondar2025}, in the kite design combining Josephson junctions and inductors~\cite{brooks2013,smith2020,smith2022,smith2025,roverch2026}, in hybrid superconducting devices \cite{feinberg2015,larsen2020parity-protected,schrade2022,messelot2024phase,arnault2025multiplet,leblanc2025,leblanc2025gate,banszerus2024voltage-controlled,ciaccia2024charge-4e,messelot2024,banszerus2025hybrid}. Alternatively, the first harmonic can be suppressed by symmetry, as in twisted $d$-wave junctions \cite{brosco2024superconducting,coppo2024,confalone2025}. More generally, $\cos(2\varphi)$ elements open novel routes towards quantum computation with superconducting circuits \cite{chirolli2021enhanced,chirolli2022swap,maiani2022,chirolli2026fraxonium}, and they have been recently utilized in the implementation of a superconducting grid-state qubit~\cite{nguyen2025}. 

The main difficulty of interference-based realizations is that they require a precise balance between the two arms of the SQUID and are  sensitive to flux noise around $\Phi_0/2$.  Alternative strategies have been explored in hybrid systems with interacting quantum dots, both at equilibrium with attractive interactions \cite{chirolli2024} and out of equilibrium with repulsive interactions \cite{chirolli2026}, where a purely many-body quartet resonance emerges. Quartet currents have also been proposed \cite{cuevas2007voltage-induced,freyn2011production,jonckheere2013multipair,jacquet2020} and observed \cite{pfeffer2014subgap,cohen2018nonlocal,huang2022evidence} in three-terminal Josephson junctions, but only under finite voltage bias. It would therefore be highly desirable to have a device that relies neither on interference  nor on nonequilibrium conditions, and that exploits the naturally repulsive Coulomb interaction of quantum dots.

\begin{figure}[t!]
    \centering
    \includegraphics[width=0.99\linewidth]{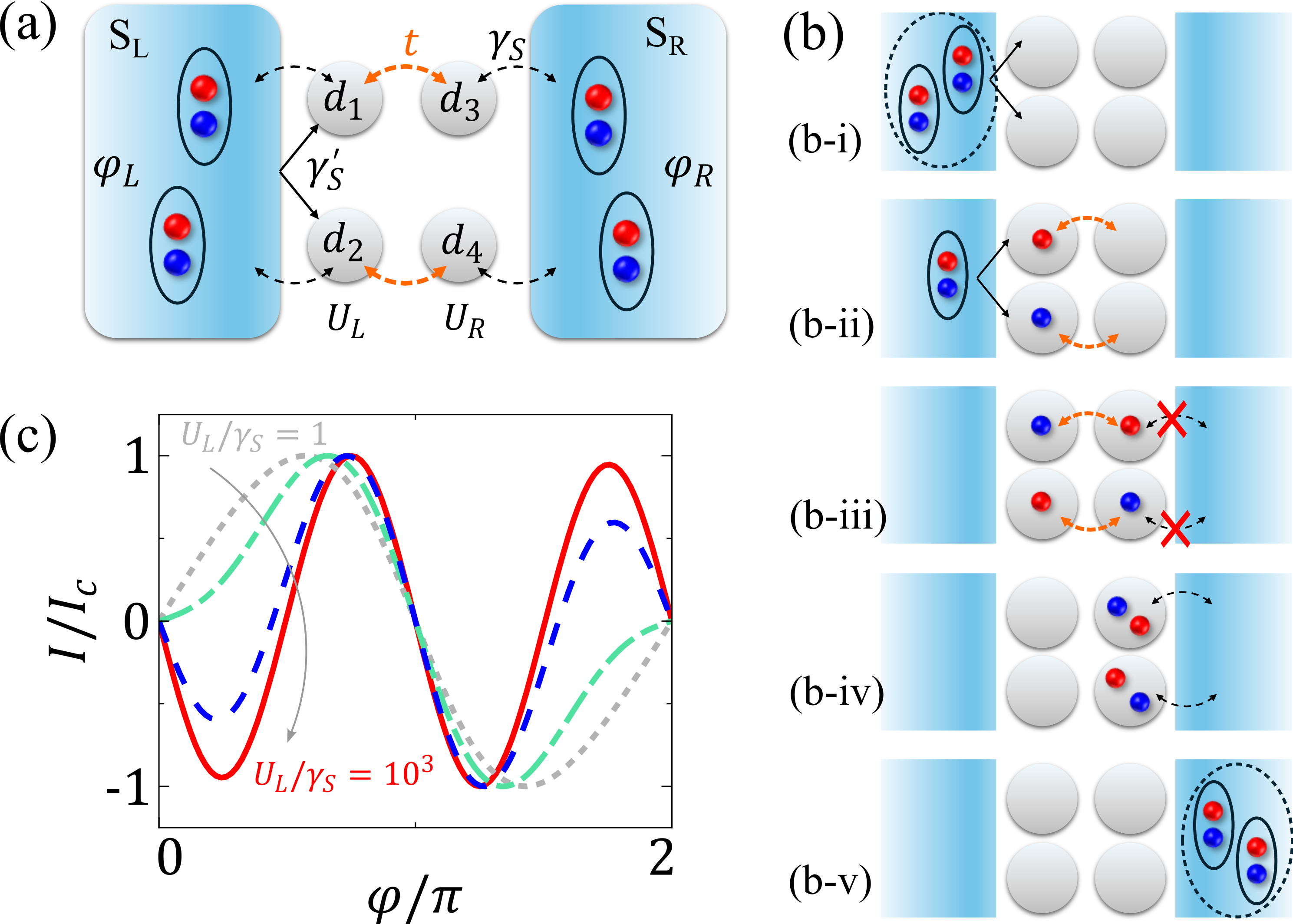}
    \caption{
    (a) Schematic of the four-quantum-dot Josephson junction.
    Local Andreev reflection with amplitude $\gamma_S$ and crossed Andreev reflection with amplitude $\gamma^{\prime}_S$ are denoted by the black dashed and black solid arrows, respectively.
    The orange dashed arrows denote single-electron hopping with strength $t$, coupling $d_1$ ($d_2$) to $d_3$ ($d_4$).
    The superconducting phases are denoted by $\varphi_L$ and $\varphi_R$, while $U_L$ ($U_R$) denotes the intradot interaction in the left (right) pair of quantum dots.
    (b) Schematic transport cycle for the double Cooper pair transfer in the limit of large $U_L$.
    (c) Normalized current-phase relation $I/I_c$ for $U_L/\gamma_S=1$ (dotted gray), $U_L/\gamma_S=10$ (dash-dotted green), $U_L/\gamma_S=100$ (dashed blue), and $U_L/\gamma_S=1000$ (solid red).
    Other parameters: $\gamma_S=1$, $\gamma^{\prime}_S=1$, $U_R=1$, and $t_{13}=t_{24}=t=1$; $|\Delta|\to\infty$ limit is implemented.
    }
    \label{figure1}
\end{figure}

In this work, we propose a hybrid superconducting device showing a Josephson energy-phase relation dominated by the two-Cooper-pair component. The basic mechanism we exploit is Cooper-pair splitting by two interacting quantum dots, as first proposed in the seminal work \cite{recher2001}. There, the strong charging energy $U$ of the quantum dots effectively inhibits local Andreev reflection (LAR), leaving only crossed Andreev reflection (CAR) through the dots into normal drain leads. Since then, Cooper-pair splitting in double quantum dots has been investigated theoretically  \cite{recher2002,recher2003,sauret2004quantum,chevallier2011current,burset2011microscopic,hussein2016,brange2021dynamic,brange2024adiabatic,delfino2025cooper-pair,PhysRevB.102.075430}, and demonstrated in with experiments 
\cite{russo2005experimental,hofstetter2009cooper,hofstetter2011finite-bias,schindele2012near-unity,das2012high-efficiency,fulop2014local,tan2015cooper,fulop2015magnetic,borzenets2016high,baba2018cooper-pair,pandey2021ballistic,ranni2021real-time,scherubl2022from,kurtossy2022parallel,bordoloi2022spin,wang2022singlet,dejong2023controllable,bordin2024crossed,wang2023triplet}. 
If normal drain leads are replaced by a superconducting one \cite{choi2000,defranceschi2010,martin-rodero2011} and the geometry of the devices is arranged to suppress nonlocal recombination (CAR) of the split pair, each electron must pay an energy $\Delta$ to tunnel individually into the drain, and the Josephson current is suppressed in the limit $U,|\Delta|\to\infty$.

\begin{figure}
    \centering
    \includegraphics[width=0.99\linewidth]{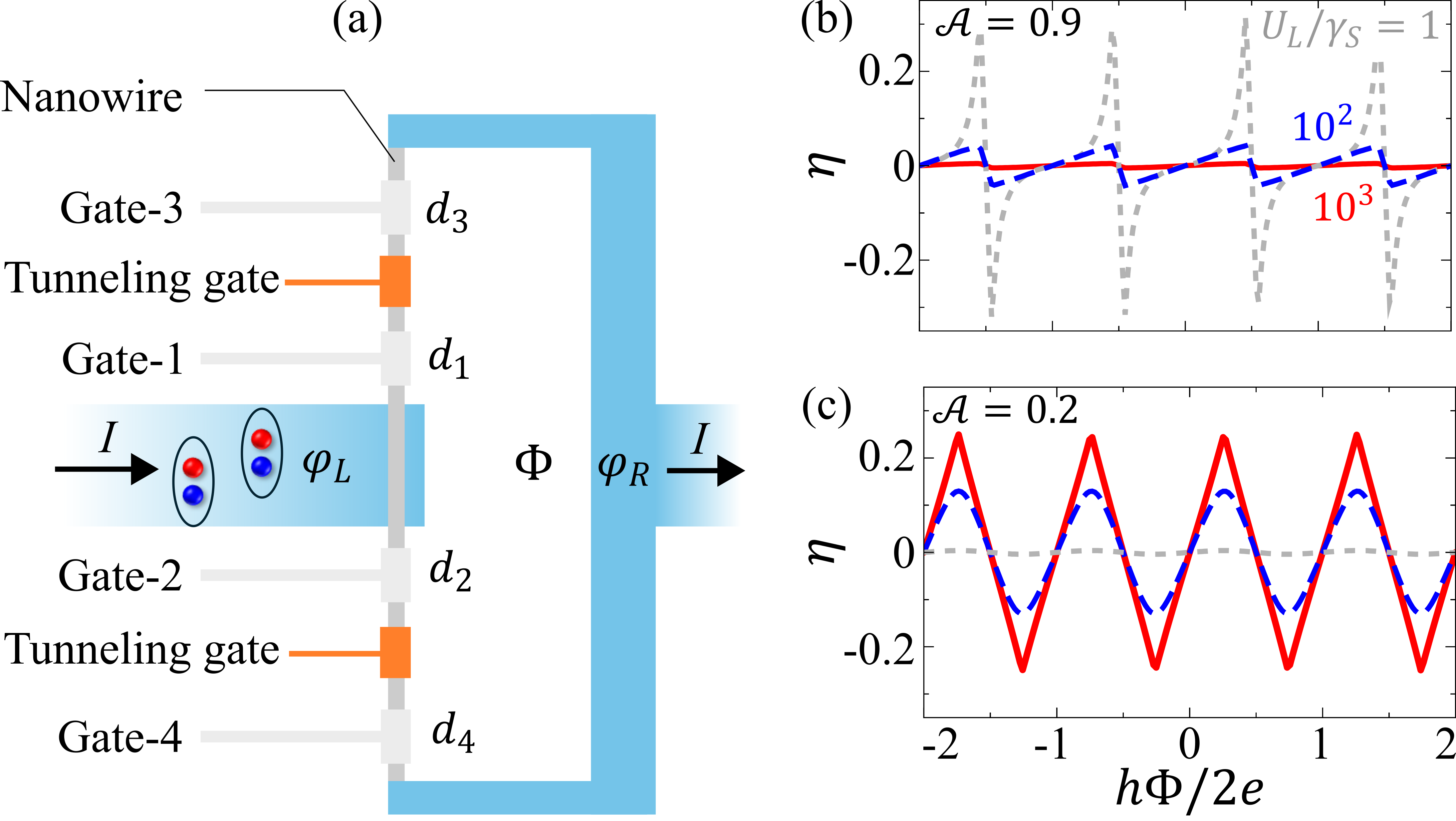}
\caption{
(a) Schematic SQUID setup for detecting the two-Cooper-pair current via the four-quantum-dot Josephson junction [Fig.\,\ref{figure1}], subjected to an external magnetic flux $\Phi$. The quantum dots ($d_i$) are defined in a nanowire (dark gray) by an electrostatic gate (light gray), while the interdot tunneling is controlled by tunneling gates (orange).
(b) Josephson diode efficiency of the SQUID in (a) as a function of the magnetic flux $\Phi$ for a coupling asymmetry $\mathcal{A}=0.9$ and different intradot Hubbard interactions: cases with $U_L/\gamma_S=1$, $100$ and $1000$ are denoted by gray dotted, blue dashed and red lines, respectively.
(c) Same as (b), but for $\mathcal{A}=0.2$.
}
\label{figure2}
\end{figure}

The idea of the present work is to modify the setup and introduce a buffer between the Cooper-pair splitter and the  superconducting drain, as shown in Fig.~\ref{figure1}(a). The buffer is represented by two additional weakly interacting dots, which can temporarily host the split pair and allow a second pair to split, thus unblocking two-Cooper-pair transfer without paying the large charging energy cost. Using perturbation theory in the tunneling amplitude we show that the two-Cooper pair process is the only one that survives the limit $U,|\Delta|\to\infty$, and numerical results confirm its robustness to asymmetries in the  interactions strengths, tunneling amplitudes, and dot levels. 
A clear signature of the two-Cooper pair transport can be seen in dc-transport measurements, by closing the drain superconducting lead on a loop enclosing a flux $\Phi$, as shown in Fig.~\ref{figure2}(a). Analysis of the diode effect \cite{souto2022,leblanc2025,nadeem2023,chirolli2025,79tj-c3y4,PhysRevApplied.21.054057} reveals a regime that can be ascribed to dominance of the two-Cooper pair current over the conventional contribution of individual Cooper pairs, allowing indirect detection of the phenomenon.

To the best of our knowledge, Cooper-pair splitting has never been exploited to realize an equilibrium Josephson junction dominated by two-Cooper-pair transport without the aid of an external flux. Our results open the way to gate-tunable, equilibrium $\cos(2\varphi)$ junctions as building blocks for parity-protected qubits. Arrays of such junctions could also serve as a controllable platform for charge-$4e$ superconducting correlations.

{\it Hamiltonian and setup.---}
We consider the system  schematically shown in Fig.~\ref{figure1}(a). Two superconductors are connected through four quantum dots. Dots 1 and 2 are coupled to the left superconductor $\mathrm{S}_L$ through LAR and CAR processes and form the Cooper-pair splitter, whereas dots 3 and 4 are coupled to the right superconductor $\mathrm{S}_R$ only through LAR processes and form the buffer. 
This condition can be easily arranged by proper design of the circuit, as shown in Fig.~\ref{figure2}(a). 
In addition, conventional single-electron hopping connects dots 1 to 3 and dots 2 to 4, without cross coupling. 
The system is described by the Hamiltonian
\begin{equation}
H_{\rm JJ}=H_{\rm S}+H_{\rm d}+
T
\label{Hjj}
\end{equation}
where $H_{\rm S}=\sum_{{\bm k}\sigma l }\xi_{{\bm k}l}c^\dag_{{\bm k}\sigma l}c_{{\bm k}\sigma l}+\sum_{l{\bm k}}|\Delta_l|c^\dag_{{\bm k}\uparrow l}c^\dag_{-{\bm k}\downarrow l}+{\rm H.c.}$ describes the superconducting leads $l=L,R$, characterized by a gap $|\Delta_l|$ and single-particle energy levels $\xi_{{\bm k}l}$, which we assume equal for the two leads.
The superconducting phase of the leads has been gauged to the tunneling Hamiltonian $T$. The quantum dots are described by
\begin{equation}
H_{\rm d}=\sum_{i,\sigma}\epsilon_{i}n_{i\sigma}+\sum_{i}U_{i}n_{i,\uparrow}n_{i,\downarrow}
,
\label{Hd}
\end{equation}
where $\epsilon_{i}$ denotes the gate-tunable single-electron energy level of dot $i$, and we assume $\epsilon_{i}=\epsilon$ for all four dots. Here, 
$n_i=n_{i,\uparrow}+n_{i,\downarrow}$ and
$n_{i,\sigma}=d_{i,\sigma}^{\dagger}d_{i,\sigma}$,
with $d_{i,\sigma}$ the fermionic annihilation operator for an electron with spin $\sigma$ on dot $i$.
We assume identical intradot Coulomb interactions within each pair of quantum dots,
$U_1=U_2\equiv U_{\rm L}$ and $U_3=U_4\equiv U_{\rm R}$, and consider the regime
\begin{equation}
    U_{\rm L}\gg U_{\rm R},
    \label{ULggUR}
\end{equation}
such that strong Coulomb repulsion $U_{\rm L}$ suppresses double occupation of the Cooper-pair splitter dots, whereas weaker Coulomb repulsion $U_{\rm R}$ allows the buffer dots to accommodate two electrons. 

The coupling between the quantum dots and the superconductors, as well as the tunneling between the quantum dots, is described by $T$, which can be decomposed as
\begin{equation}
T=T_{\rm L}+T_{\rm R}+T_{\rm D},
\label{HT}
\end{equation}
where 
\begin{equation}
T_{l}=-\sum_{j\sigma}t_{lj}e^{i\varphi_{l}/2} d^{\dagger}_{j,\sigma}\psi_{l,\sigma}(\bm{r}_l)+{\rm H.c.},
\label{Hl}
\end{equation}
with $t_{lj}$ the tunneling amplitude between dot $j$ and the point $\bm{r}_{l}$ in superconductor $l$, which has superconducting phase $\varphi_{l}$.
Here, $\psi_{l,\sigma}(\bm{r})$ is the electron field operator with spin $\sigma$ at position $\bm{r}$ in superconductor $l$, related to $c_{\bm{k}\sigma l}$ by 
$\psi_{l,\sigma}(\bm{r})=1/\sqrt{V}\sum_{{\bm k}}e^{i\bm{k}\cdot\bm{r}}c_{\bm{k}\sigma l}$
where $V$ is the volume of each superconductors, assumed to be equal for both. The interdot tunneling in Eq.\,(\ref{HT}) is 
\begin{equation}
T_{\rm D}=-\sum_{\sigma}t_{13}d^{\dagger}_{1,\sigma}d_{3,\sigma}+t_{24}d^{\dagger}_{2,\sigma}d_{4,\sigma}+{\rm H.c.},
\label{HD}
\end{equation}
where $t_{13}$ ($t_{24}$) denotes the hopping amplitude between  dot 1 (2) and dot 3 (4). 
Hereafter, we assume all the tunneling processes are spin-preserving and set for simplicity 
$t_{\rm{L}1}=t_{\rm{L}2}=t_{\rm L}$ and $t_{\rm{R}3}=t_{\rm{R}4}=t_{\rm{R}}$ in Eq.\,(\ref{Hl}). 
In addition, we first take the two interdot hopping amplitudes in Eq.\,(\ref{HD}) to be identical, $t_{13}=t_{24}=t$, while cases with asymmetric interdot tunneling ($t_{13}\neq t_{24}$) will be discussed later.

\textit{Josephson effect.---}
We now inspect the Josephson effect of the four-quantum-dot junction described by Eq.\,(\ref{Hjj}).
We employ perturbation theory in the tunneling amplitudes within the $T$-matrix formalism \cite{recher2001,rozhkov2001,recher2003,PhysRevB.81.144519}, taking the unperturbed Hamiltonian as $H_{0}=H_{\rm S}+H_{\rm d}$ [see Eqs.\,(\ref{Hjj})]. The initial and final state $|\Psi\rangle=|0\rangle|\Psi_L\rangle|\Psi_R\rangle$ is given by the product of the BCS ground states $|\Psi_l\rangle$ and the vacuum state $|0\rangle$ of the four-dot system, and the Josephson energy is given as a series expansion in the tunneling matrix elements
\begin{equation}
    E_J(\varphi)= \sum_{n=0}^\infty \langle T(G_0T)^n\rangle,
\end{equation}
where $\varphi = \varphi_{\rm R} - \varphi_{\rm L}$ is the phase difference between the two superconducting leads and $G_0=(E+i0^+-H_0)^{-1}$ evaluated at $E=0$. The expansion gives rise to a sum of terms, each describing the coherent transfer of $n$ Cooper pairs and contributing a term $\propto\cos(n\varphi)$ to the Josephson energy (see \cite{SM} for details).
Retaining the contributions from the coherent transfer of one and two Cooper pairs in the limit $\Delta\to\infty$, we obtain the Josephson energy  \cite{SM}
\begin{eqnarray}
\label{EJ}
E^{\rm J}(\varphi)=
-\Gamma_{\rm 2e}\cos\varphi
+
\Gamma_{\rm 4e}\cos2\varphi,
\end{eqnarray}
with the corresponding Josephson coupling strengths 
\begin{eqnarray}
\Gamma_{\rm 2e}
&=&
\frac{t^{2}\gamma_{L}\gamma_{R}}{4\epsilon\left(2\epsilon+U_{R}\right)\left(2\epsilon+U_{L}\right)},
\label{Gamma2e}
\\
\Gamma_{\rm 4e}
&=&
\frac{t^{4}\gamma^{\prime2}_{L}\gamma^{2}_{R}}{32(2\epsilon)^{4}(2\epsilon+U_{R})^{2}\left(4\epsilon+U_{R}\right)}
\label{Gamma4e}
.
\end{eqnarray}
where  $\gamma_{l}=2\pi\rho_{l}|t_{l}|^{2}$ is the bare tunneling rate between the $S_l$ and one of the two adjacent dots, with $\rho_l$ the normal-state density of states per spin of superconducting lead $l$ at the Fermi energy. The rate $\gamma_l$ controls local Andreev reflection (LAR) from a superconducting lead and an individual quantum dot. In turn, the rate $\gamma^{\prime}_{l}
=\gamma_{l}
\frac{\sin(k_{F}\delta r_{ij})}{k_{F}\delta r_{ij}}
e^{-\delta r_{ij}/\pi l_{\rm sc}}$
describes crossed Andreev reflection (CAR) from a superconducting lead to  dots $i$ and $j$ separated by a distance  
$\delta r_{ij}$,  
with $k_{F}$ the Fermi momentum, and $l_{\rm sc}$ the superconducting coherence length.

Our key assumption on the coupling of the buffer dots 3 and 4 to $S_R$ [see Figs.~\ref{figure1}(a) and \ref{figure2}(a)] is $\gamma_R'=0$, so that a Cooper pair cannot split at $S_L$ and recombine at $S_R$. As a consequence, $\Gamma_{2e}$ contains only the local couplings $\gamma_l$, which control LAR, and the single-Cooper-pair contribution to the current is the coherent sum of the paths through the upper (1--3) and lower (2--4) branches. Since this process requires double occupancy of dots 1 and 2, it is controlled by their charging energy and is suppressed in the limit $U_L\to\infty$ \cite{recher2001}.

On the other hand, $\Gamma_{4e}$ originates from a multiplicity of microscopic processes summed coherently \cite{SM}. All but one of them are suppressed: those whose intermediate states contain quasiparticles in the superconductors vanish for $|\Delta|\to\infty$, and those in which Cooper pairs enter the Cooper-pair splitter via LAR vanish for $U_L\to\infty$. The only surviving process is the sequential splitting of two Cooper pairs through the buffer, illustrated in Fig.~\ref{figure1}(b). It consists of the following steps: (i) a first Cooper pair from $S_L$ is split between dots 1 and 2; (ii) the resulting singlet is transferred to dot 3 and 4 as a singlet; (iii) a second Cooper pair from $S_L$ is split to dots 1 and 2; (iv) the second pair is transferred to dots 3 and 4, which become dobly occupied; (v) the two local pairs in dots 3 and 4 are transferred to $S_R$ via LAR. Conversely, a Cooper pair entering dot 3 (or 4) from $S_R$ via LAR cannot recombine in $S_L$ on its own: it has to wait for a second pair on the other buffer dot, so that two successive recombinations can take place through the splitter. 
The resulting Josephson current-phase relation (CPR) $I^J=(2e/\hbar)\partial E^J/\partial \varphi$ reads
\begin{equation}
\lim_{|\Delta|,U_{\rm L}\rightarrow\infty}I^{\rm J}
=
-\frac{4e}{\hbar}
\Gamma_{\rm 4e}
\sin2\varphi,
\label{I4e}
\end{equation}
which is $\pi$-periodic in the phase difference $\varphi$. This is the most important result of the present work.  
Importantly, the perturbative calculation (see \cite{SM}) also allows us to estimate the correction due to finite $|\Delta|$ in the limit $U_{\rm L}\to\infty$, which is a more relevant experimental configuration, $\delta I^J = \frac{2e}{\hbar}
\frac{t^{2}}{\pi|\Delta|(2\epsilon)^{2}}
\gamma_{L}\gamma_{R}
\sin(\varphi)+{\cal O}(1/|\Delta|^2)$
where we assumed $0<\epsilon\ll |\Delta|$, and for $U_{\rm R}=0$, we have  $\Gamma_{4e}/\Gamma_{2e}= \pi t^2 (\gamma_L'^2/\gamma_{\rm L}) \gamma_R |\Delta|/(2^{11}\epsilon^5)$.

{\it $|\Delta|\to\infty$ limit.---} In the limit $|\Delta|\rightarrow\infty$ quasiparticle excitations in the superconducting leads become inaccessible and the scattering between leads and dots becomes elastic \cite{rohzkov-arovas2000,rozhkov2001}. The net effect on the dots is the emergence of local and nonlocal proximity pairing amplitudes, so that the Hamiltonian becomes  
$H^{\prime}_{\rm JJ}
=H_{\rm d}+H_{\rm Sd}+T_{\rm D}$, 
where $H_{\rm Sd}$ is given by
\begin{equation}
H_{\rm {Sd}}=
-\sum_{i=1}^{4}\frac{\gamma_i}{2}
d_{i,\uparrow}^{\dagger}d_{i,\downarrow}^{\dagger}
-\frac{\gamma'_{\rm L}}{2}
\left(
d_{1,\uparrow}^{\dagger}d_{2,\downarrow}^{\dagger}
-d_{1,\downarrow}^{\dagger}d_{2,\uparrow}^{\dagger}
\right),
\label{HSd}
\end{equation}
where $\gamma_1=\gamma_2=\gamma_{\rm L},
\quad
\gamma_3=\gamma_4=\gamma_{\rm R}e^{i\varphi}$. 
At zero temperature, the Josephson current can be evaluated numerically as $I^{\rm J}(\varphi)
=
\frac{2e}{\hbar}
\langle\psi_{\rm gs}|
\partial_{\varphi}H_{\rm Sd}
|\psi_{\rm gs}\rangle,
\label{currentlargeDelta}
$ where $|\psi_{\rm gs}\rangle$ denotes the ground state of $H^{\prime}_{\rm JJ}$. 
Setting $\gamma_{\rm L}=\gamma_{\rm R}=\gamma_{\rm S}$ as the energy unit, Fig.\,\ref{figure1} (c) shows the evolution of $I^{\rm J}(\varphi)/I_c$ versus $U_{\rm L}/\gamma_S$, with $I_c=\max_{\varphi}I(\varphi)$ the positive critical current. The result can be understood via the expresssion Eq.\,(\ref{EJ}):
For $U_{\rm L}=U_{\rm R}=\gamma_{\rm S}$, the CPR approximately follows the conventional sinusoidal dependence $I(\varphi)\sim\sin\varphi$, indicating that Josephson transport is dominated by single-Cooper-pair tunneling, corresponding to $\Gamma_{\rm 2e}\gg\Gamma_{\rm 4e}$.
As $U_{\rm L}/\gamma_{\rm S}$ increases at fixed $U_{\rm R}/\gamma_{\rm S}$, the CPR progressively deviates from the conventional sinusoidal behavior, showing an enhanced second harmonic. In the strongly interacting regime, the first harmonic is strongly suppressed, $\Gamma_{\rm 2e}\ll\Gamma_{\rm 4e}$, and the CPR is dominated by the second Josephson harmonic, with an approximately $\pi$-periodic dependence.

\textit{Josephson diode effect (JDE).---}The emergence of a $\pi$-periodic supercurrent can be detected through the Josephson diode effect.
To this end, we propose the SQUID setup illustrated in Fig.\,\ref{figure2}(a), in which two superconducting leads are connected through a nanowire-based  
junction, with electrostatically defined dots and barriers. 
The geometry enables $\delta r_{34}\gg\delta r_{12}$, naturally suppressing CAR between the right superconductor and dots $d_3$ and $d_4$. The SQUID is threaded by an  external magnetic flux $\Phi$, which is incorporated in the interdot hopping amplitudes as Peierls phases 
$t_{13}\rightarrow t_{13}e^{-i\tilde{\Phi}/2}$
and
$t_{24}\rightarrow t_{24}e^{i\tilde{\Phi}/2}$,
where $\tilde{\Phi}=2\pi e\Phi/h$ \cite{SM}.
The Josephson current at lowest orders in the tunneling becomes \cite{SM}
\begin{equation}
I^{\rm J}(\varphi)=
\frac{2e}{\hbar}\Gamma_{\rm 2e}P_{\Phi}
\sin(\varphi-\varphi_{\Phi})
-\frac{4e}{\hbar}\mathcal{A}\Gamma_{\rm 4e}
\sin(2\varphi),
\label{IJFlux}
\end{equation}
where
$\mathcal{A}\equiv t_{24}^{2}/t_{13}^{2}$,
$P_\Phi=(1+{\cal A})\sqrt{1-\tau \sin^2(\tilde\Phi)}$ with $\tau=\frac{4{\cal A}}{(1+{\cal A})^2}$,  
and $\varphi_{\Phi}
=
\arctan\left(
\frac{1-\mathcal{A}}{1+\mathcal{A}}
\tan\tilde{\Phi}
\right)$. The latter 
denotes an anomalous phase shift which is absent in the second harmonic thanks to the setup that favors Cooper pair splitting, in which the split electrons are guided along the top or bottom paths, acquiring opposite Peierls phases, which cancel in the resulting tunneling amplitude.

The interference between the two terms at finite flux can produce a nonreciprocal Josephson effect, which is quantified by the diode efficiency, $\eta
=
\frac{I^{\rm J}_{+ c}-|I^{\rm J}_{-c}|}
{I^{\rm J}_{+c}+|I^{\rm J}_{-c}|}$, where $I^{\rm J}_{\pm c}=\max/\min_{\varphi}I^{\rm J}(\varphi)$
denote the positive and negative critical currents,
respectively.
An enhanced diode response is expected when the two harmonics interfere constructively for one current direction and destructively for the opposite direction.
The optimal-interference condition is $\varphi_{\Phi}^{c}=\pi/4+n\pi/2$, with $n\in\mathbb{Z}$, corresponding to suitable $\mathcal{A}_{c}$ and $\tilde{\Phi}_{c}$ and
leading to
\begin{equation}
\label{etamax}
\eta_{\rm max}
=
\max_{\tilde{\Phi}}
\left|
\frac{I^{\rm J}_{+c}-|I^{\rm J}_{-c}|}
{I^{\rm J}_{+c}+|I^{\rm J}_{-c}|}
\right|
=
\frac{\mathcal{A}_{c}}{P_{\Phi_c}}
\frac{2\Gamma_{\rm 4e}}{\Gamma_{\rm 2e}},
\end{equation}
where
$P_{\Phi_c}
=\sqrt{2}(1+\mathcal{A}_{c})
|\cos\tilde{\Phi}_{c}|$
and $\tilde{\Phi}_{c}\neq n\pi/2$. 
Eq.\,\eqref{etamax} highlights the role of the second harmonic in generating the Josephson diode effect. 
In particular, when $\Gamma_{\rm 4e}=0$, the CPR contains only the first harmonic and the diode efficiency vanishes.
Figs~\ref{figure2}(b) and (c) present the flux dependence of the
JDE efficiency for different $U_{\rm L}/\gamma_{\rm S}$ at
$\mathcal{A}=0.9$ and $\mathcal{A}=0.2$, respectively.
For $\mathcal{A}=0.9$, corresponding to weak tunneling asymmetry, a pronounced JDE is observed only at relatively small $U_{\rm L}/\gamma_{\rm S}$. In contrast, for $\mathcal{A}=0.2$, corresponding to a strong tunneling asymmetry, the diode efficiency increases at large $U_{\rm L}/\gamma_{\rm S}$. The two behaviors are associated with the interplay between the two harmonics, and they can be used for indirect diagnostic purposes.

Specifically, the opposite behaviors are understood as follows: for weak asymmetry (${\cal A}\lesssim 1$) and in the vicinity of $\tilde{\Phi}=\pi/2+n\pi$,   destructive interference  strongly suppresses the first-harmonic amplitude ($P_{\Phi} \ll 1$), making it comparable to second-harmonic one, thus enhancing the interplay between the two harmonics and, consequently, the JDE efficiency. Conversely, as $U_{\rm L}/\gamma_{\rm S}$ is further increased, the strong Coulomb interaction progressively suppresses the first harmonic (see Eq.,(\ref{Gamma2e})), leaving the second one dominant and thus reducing the diode efficiency. On the other hand, for large asymmetry (${\cal A}\ll 1$), the diode effect is negligible at small $U_{\rm L}/\gamma_{\rm S}$, since the second harmonic is much weaker than the first. Increasing $U_{\rm L}/\gamma_{\rm S}$ selectively suppresses
the first harmonic until the two become comparable,
$\Gamma_{\rm 2e}P_{\Phi}
\sim2\mathcal{A}\Gamma_{\rm 4e}$, and their interplay gives rise to a pronounced Josephson diode effect. 
Notably, in this strongly asymmetric regime, the diode effect persists over a broad range of magnetic flux. It vanishes only at $\tilde{\Phi}=n\pi/2$ ($n\in\mathbb{Z}$), where $\sin(2\varphi_{\Phi})=0$: there the relative phase between the two harmonics is $0$ or $\pi$, and the critical currents become symmetric again due to restored time-reversal symmetry. Conversely, the diode efficiency is largest near $\tilde{\Phi}=\pi/4+n\pi/2$, where $|\sin(2\varphi_{\Phi})|=1$.

These results demonstrate that the intradot Coulomb interaction controls the relative strengths of the single- and two-Cooper-pair currents, while the magnetic flux and tunneling asymmetry control their interference. In particular, the tunneling asymmetry can be controlled with an external gate, so that the phenomenology outlined can be employed to experimentally extract the dominant character of the junction, thus providing an indirect signature of coherent two-Cooper-pair transport.

\textit{Conclusion.---}We have proposed a superconducting hybrid device in which a Cooper-pair splitter, combined with a buffer dot, efficiently splits two consecutive Cooper pairs, while charging effects and the large superconducting gap suppress all competing processes. The device thus enables two-Cooper-pair transport without requiring either a magnetic flux or nonequilibrium conditions, and realizes an $h/4e$ Josephson effect. Its simplicity makes it well suited for experimental realization with ingredients readily available in current solid-state laboratories.

\textit{Acknowledgments.---}P.-H. F. and  L. C. acknowledge financial support from the Fondazione Cariplo under the grant 2023-2594.


%

\end{document}


\title{Supplementary Information for ``Two-Cooper-pair Josephson junction from Cooper-pair splitting"}

 \author{Pei-Hao Fu}
\email{phy.phfu@gmail.com}
\affiliation{Department of Physics and Astronomy, University of Florence, I-50019 Sesto Fiorentino, Italy}

\author{Luca Chirolli}
\email{luca.chirolli@unifi.it}
\affiliation{Department of Physics and Astronomy, University of Florence, I-50019 Sesto Fiorentino, Italy}
 
\date{\today}
\maketitle
\tableofcontents
\newpage

\section{Setup and Hamiltonian}

The Josephson junction shown in Fig.\,1 in the main text consists of two superconductors connected through a four-quantum-dot system; see also Fig.\,\ref{figureS1}(a).
Quantum dots $1$ and $2$ are coupled to the left superconductor, whereas quantum dots $3$ and $4$ are coupled to the right superconductor. 
The setup is described by the Hamiltonian
\begin{equation}
H_{\rm JJ}=H_{\rm S}+H_{\rm d}+ T .
\label{Hjj}
\end{equation}
Here,
\begin{equation}
H_{\rm S}=\sum_{l,\bm{k},\sigma}E_{l,\bm{k}}
\gamma^{\dagger}_{l,\bm{k},\sigma}
\gamma_{l,\bm{k},\sigma},
\label{HS}
\end{equation}
describes the BCS superconducting leads, where $l\in\{\rm L,\rm R\}$ labels the left and right superconductors and $\sigma=\uparrow,\downarrow$ denotes the spin. 
The quasiparticle dispersion is given by
$E_{l,\bm{k}}=\sqrt{\xi^{2}_{l,\bm{k}}+|\Delta_{l}|^2}$, where
$\xi_{l,\bm{k}}=\epsilon_{l,\bm{k}}-\mu_{l,\rm S}$ is the normal-state single-electron energy measured relative to the Fermi level $\mu_{l,\rm S}$ in the corresponding superconducting lead, and we assume it satisfies $\xi_{-{\bm k}}=\xi_{\bm k}$. 
The operators $\gamma_{l,\bm{k},\sigma}$ describe Bogoliubov quasiparticle excitations above the BCS ground state $|0\rangle_{l,\rm S}$, defined by
$\gamma_{l,\bm{k},\sigma}|0\rangle_{l,\rm S}=0$.
They are related to the electron annihilation and creation operators
$c_{l,\bm{k},\sigma}$ and $c^{\dagger}_{l,\bm{k},\sigma}$ through the Bogoliubov transformation 
\begin{eqnarray}
c_{l,\bm{k},\sigma} & = &
u_{l,\bm{k}}\gamma_{l,\bm{k},\sigma} 
+
\sigma v_{l,\bm{k}}
\gamma^{\dagger}_{l,-\bm{k},-\sigma},
\label{clksigma}\\
c^{\dagger}_{l,\bm{k},\sigma} & = &
u_{l,\bm{k}}\gamma^{\dagger}_{l,\bm{k},\sigma}
+
\sigma v_{l,\bm{k}}
\gamma_{l,-\bm{k},-\sigma},
\end{eqnarray}
where $\sigma=+1$ ($-1$) for $\sigma=\uparrow$ ($\downarrow$), and
\begin{eqnarray}
u_{l,\bm{k}}= 
\frac{1}{\sqrt{2}}
\sqrt{1+\frac{\xi_{l,\bm{k}}}{E_{l,\bm{k}}}},
\quad
v_{l,\bm{k}}=
\frac{1}{\sqrt{2}}
\sqrt{1-\frac{\xi_{l,\bm{k}}}{E_{l,\bm{k}}}},
\label{ulk}
\end{eqnarray}
are the usual BCS coherence factors, and we have gauged the phase of the gap on the tunneling Hamiltonian $H_{\rm{T}}$ [see Eq.\,(\ref{HT}) below].

The two superconductors described by Eq.\,(\ref{HS}) are connected through a four-quantum-dot system, whose Hamiltonian is given by the second term of Eq.\,(\ref{Hjj}) and can be written as 
\begin{equation}
H_{d}=\sum_{i}\epsilon_{i}n_{i\sigma}+\sum_{i}U_{i}n_{i,\uparrow}n_{i,\downarrow}
.
\label{Hd}
\end{equation}
Here, $\epsilon_{i}$ denotes the gate-tunable single-electron energy level of dot $i$. 
We assume that all four dots are tuned to the same level, such that $\epsilon_{i}=\epsilon$, while the other levels are neglected.
The parameter $U_{i}$ describes the intradot Hubbard interaction, which we assume to be repulsive, $U_{i}>0$.
In Eq.~(\ref{Hd}),
$n_{i}=n_{i,\uparrow}+n_{i,\downarrow}$ and
$n_{i,\sigma}=d^{\dagger}_{i,\sigma}d_{i,\sigma}$, where $d_{i,\sigma}$ is the fermionic annihilation operator for an electron with spin $\sigma$ on dot $i$.
To achieve the two-Cooper-pair junction via two subsequent Cooper pair splitting, one important regime is 
\begin{equation}
U_{\rm L}\gg U_{\rm R},
\label{ULggUR}
\end{equation}
where we have assumed identical intradot interactions within each pair of quantum dots,
$U_{1}=U_{2}\equiv U_{\rm L}$ and $U_{3}=U_{4}\equiv U_{\rm R}$, while allowing the interaction strengths of the two pairs to differ. 

The coupling between the four-quantum-dot system and the superconductors, as well as the tunneling between the quantum dots, is described by $T$ in Eq.\,(\ref{Hjj}), which can be decomposed as
\begin{equation}
T = T_{\rm L} + T_{\rm R} + T_{\rm D} ,
\label{HT}
\end{equation}
where the first four terms can be specified as
\begin{equation}
\bar{T}_{ l} =
T_{\rm{D} l } + T_{l \rm{D}}
=
-\sum_{\sigma}t_{lj}d^{\dagger}_{j,\sigma}\psi_{\sigma}(\bm{r}_{l})+h.c.,
\label{Hl}
\end{equation}
with  $l\in\{\rm L,\rm R \}$, describes spin-conserving tunneling between dot $j$ and the point $\bm{r}_{l}$ in superconductor $l$. 
In Eq.\,(\ref{Hl}), $T_{\rm{D} l }$ describes the tunneling event from the $l$-superconductor to the quantum dots, while $T_{l \rm{D}} = T_{l \rm{D}}^{\dagger}$ represents the reverses processes, where electrons is injected from the quantum dots to the $l$-superconductor;
Moreover, $\psi_{\sigma}(\bm{r}_{l})$ is the electron field operator at position $\bm{r}_{l}$ with spin $\sigma$, related to $c_{l,\bm{k},\sigma}$ via the orbital wave functions in the superconductor, assumed to be plane waves with wave vector $\bm{k}$ normalized on the volume $V$ of the superconductor (assumed to be equal in size for simplicity)
\begin{equation}
\psi_{\sigma}(\bm{r}_{l})=\frac{1}{\sqrt{V}}\sum_{\bm{k}}e^{i\bm{k}\cdot\bm{r}_{l}}c_{l,\bm{k},\sigma}.
\label{psiFT}
\end{equation}
The tunneling amplitude is written as
$t_{lj}=|t_{lj}|e^{i\varphi_{l}/2}$, where $\varphi_{l}$ denotes the macroscopic superconducting phase of lead $l$.
Furthermore, the last two terms in Eq.\,(\ref{HT}) are the interdot tunneling term given by
\begin{equation}
T_{\rm D}
=
T_{d_r d_l} + T_{d_l d_r}
=
-\sum_{\sigma}t_{13}d^{\dagger}_{3,\sigma}d_{1,\sigma}+t_{24}d^{\dagger}_{4,\sigma}d_{2,\sigma}+h.c.,
\label{HD}
\end{equation}
where 
$T_{d_r d_l}$ denotes the tunneling from the left two dots (dot 1 and dot 2) to  to the right two dots (dot 3 and dot 4) while 
$T_{d_l d_r} = T^{\dagger}_{d_r d_l}$ is the reverse processes; 
Here, $t_{13}$ ($t_{24}$) denotes the spin-preserving hopping amplitude between the spin-$\sigma$ state on dot 1 (2) and dot 3 (4). 
We retain only hopping processes between quantum dots aligned along the transport direction of the Josephson junction, as indicated by the orange dashed arrows in Fig.~1(a) in the main text, and neglect all other interdot hopping processes.
Hereafter, we assume equal tunneling amplitudes between each superconductor and the two quantum dots on the corresponding side of the junction, namely
$t_{\rm{L}1}=t_{\rm{L}2}=t_{\rm L}$ and $t_{\rm{R}3}=t_{\rm{R}4}=t_{\rm{R}}$ in Eq.\,(\ref{Hl}), with
$t_{l}=|t_{l}|e^{i\varphi_{l}/2}$. We also take the two interdot hopping amplitudes in Eq.\,(\ref{HD}) to be identical, $t_{13}=t_{24}=t$, and set $t_{12}=t_{34}=0$ without loss of generality.

\section{Josephson energy}

In the following, we provide a microscopic description of the Josephson effect in the four-quantum-dot junction described by Eq.~(\ref{Hjj}), arising from the transfer of Cooper pairs between the two superconductors mediated by the quantum dots. We consider Cooper-pair transport in the Coulomb-blockade regime, at zero bias voltage and neglecting thermal excitations. Within the Coulomb-blockade diamonds, sequential tunneling is suppressed and transport proceeds through cotunneling processes.

The lowest-order process transferring a single Cooper pair between the two superconducting leads involves six tunneling events. 
Such processes are supposed to provide the dominant contribution to the Josephson coupling for sufficiently small normal-state conductances. 
In addition, we consider higher-order processes involving the coherent transfer of two Cooper pairs, corresponding to four electrons, between the superconductors. 
Accordingly, we employ sixth- and twelfth-order perturbation theory in the tunneling amplitudes to obtain the corresponding contributions to the Josephson energy as the perturbation correction to the ground-state energy.

We take the unperturbed Hamiltonian to be
\begin{equation}
H_{0}=H_{\rm S}+H_{\rm d},
\label{H0}
\end{equation}
whose ground states are degenerate with respect to the number of Cooper pairs in the superconducting leads. 
The corresponding states can be written as a direct product \cite{recher2001,rozhkov2001,recher2003,PhysRevB.81.144519},
\begin{equation}
|\Psi\rangle=|0\rangle|\Psi_{\rm{L}}\rangle|\Psi_{\rm{R}}\rangle,
\label{groundstate}
\end{equation}
where $|\Psi_{l}\rangle$ denotes the state of superconducting lead $l\in\{L,R\}$ with electron number $N_{l}$ (a state with fixed number of electron in a superconducting lead is correct in the many-body treatment of the pairing interaction and we use it to keep track of the number of electrons transferred between the leads, although for calculations we use the BCS ground state expression, that does not conserve the number of particles), while $|0\rangle$ denotes the ground state of the four-quantum-dot system.
The tunneling Hamiltonian $T$ [Eq.~\eqref{HT}] couples the degenerate states with different distributions of Cooper pairs between the two superconducting leads. 
The resulting phase-dependent correction to the ground-state energy gives rise to the Josephson energy. 
In the following, we separately evaluate the contributions associated with the transfer of one Cooper pair, involving two electrons, and two Cooper pairs, involving four electrons.

\subsection{Single Cooper-pair contribution}

The contribution to the Josephson energy associated with the transfer of a single Cooper pair between the superconductors is given by \cite{PhysRevB.81.144519}
\begin{eqnarray}
E^{\rm J}_{\rm 2e}\left(\varphi\right) & = & \langle-2,2|\mathcal{T}^{\left(6\right)}|0,0\rangle+\langle2,-2|\mathcal{T}^{\left(6\right)}|0,0\rangle
\label{EJ2e}\\
 & = & 2\mathrm{Re}[\langle-2,2|\mathcal{T}^{\left(6\right)}|0,0\rangle],\nonumber 
\end{eqnarray}
where the first (second) term in the first line describes the transfer of two electrons from the left (right) superconducting lead to the right (left) superconducting lead. The operator $\mathcal{T}^{\left(6\right)}$ represents the sixth-order contribution in the tunneling Hamiltonian,
\begin{equation}
\mathcal{T}^{\left(6\right)}=T\left(G_{0}T\right)^{5},
\label{T6}
\end{equation}
where $G_{0}=\left(E-H_{0}\right)^{-1}$ is the Green's function associated with the unperturbed system [Eq.\,(\ref{H0})] evaluated at the ground-state energy ($E=0$).
We always assume the energy of the states with finite number of electrons to be larger than zero and do not consider resonant tunneling. 
This way, the denominators in the perturbation scheme are always finite. 
Each tunneling operator $T$ [Eq.\,(\ref{HT})] describes either the transfer of one electron between a quantum dot and one of the superconducting leads through $T_{ \rm{D} l}$ and $T_{  l \rm{D}}$ [Eq.\,(\ref{Hl})] or the transfer of one electron between quantum dots through $T_{d_r d_l}$ [Eq.\,(\ref{HD})]. 
A sixth-order process therefore contains six tunneling events and five intermediate virtual states, whose energies enter through the five Green's functions $G_{0}$.

At the microscopic level, permutations of the six tunneling events associated with Eq.\,(\ref{T6}) give $6!=720$ possible sequences. 
However, only a restricted subset can connect the initial and final ground states involved in the transfer of a Cooper pair. 
To identify these processes, it is useful to classify each tunneling event according to whether it is generated by $T_{ \rm{DL} }$, $T_{ \rm{RD} }$, or $T_{d_r d_l}$ given by Eqs.\,(\ref{HT})-(\ref{HD}). 
For the transfer of a Cooper pair from the left to the right superconductor, two electrons must each undergo the sequence of processes that transfers them from the left superconductor to the left dots, across the junction, and finally into the right superconductor. 
Consequently, the allowed operator sequences satisfy the following conditions:

1. The numbers of tunneling events generated by $T_{ \rm{DL} }$, $T_{ \rm{RD} }$, or $T_{d_r d_l}$ are equal, so that all electrons ejected from one superconductor are eventually injected to the other one.

2. Since operators act from right to left, $\mathcal{T}^{\left(6\right)}$ must begin with $T_{ \rm{DL} }$ (tunneling from left superconductor to the dots) and terminate with $T_{ \rm{RD} }$ (tunneling from the dots to the right superconductor) in chronological order, which leads to
\begin{equation}
T^{\left(6\right)}
=
T_{ \rm{RD} } \left(G_{0} T \right)^{4}G_{0} T_{ \rm{DL} }.
\label{T6}
\end{equation}
representing all the possible tunneling events in transferring two electrons from the left superconductor to the right one.

3. At every intermediate stage of the tunneling sequence, the number of events generated by $T_{ \rm{DL} }$ cannot be smaller than that generated by $T_{ \rm{DD} }$, which in turn cannot be smaller than that generated by $T_{ \rm{DR} }$. 
This condition ensures that an electron can tunnel between the two sides only after it has entered the quantum-dot system, and can enter the right superconductor only after it has reached the right quantum dots.

4. The number of tunneling events involving each superconducting lead must be \textit{even} in order for the superconductors to return to their BCS ground states. 
In particular, a single-Cooper-pair transfer contains two $T_{ \rm{DL} }$ and two $T_{ \rm{DR} }$ events.

\begin{figure}
    \centering
    \includegraphics[width=0.99\linewidth]{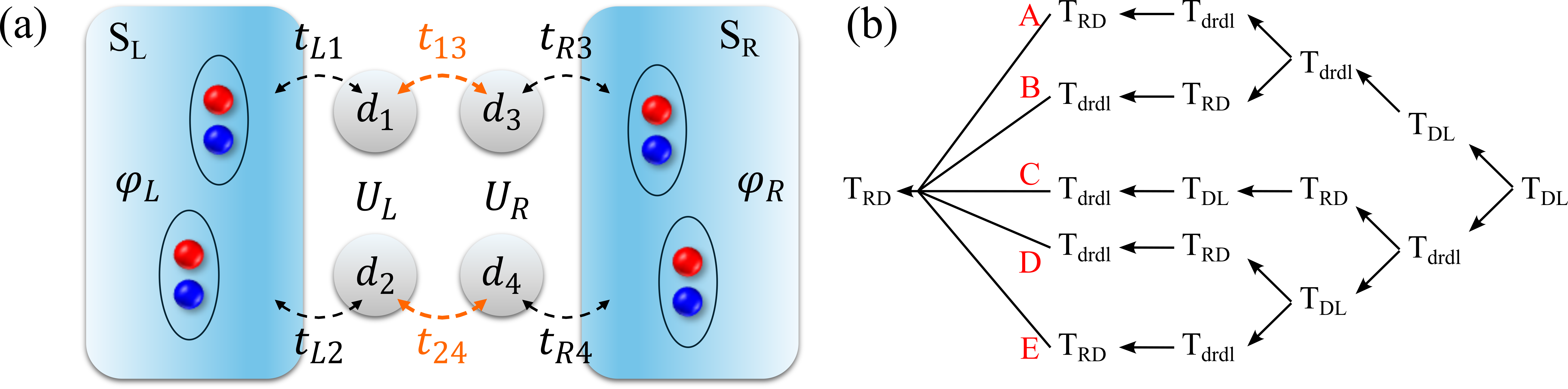}
    \caption{
    (a) Schematic Josephson junction mediated by the four-quantum-dot system associated with the Hamiltonian given by Eq.\,(\ref{Hjj}).
    (b) Schematic representation of the five allowed sixth-order tunneling sequences contributing to the transfer of a single Cooper pair from the left to the right superconductor. 
    The notation $T_{\rm{DL}}$ ($T_{\rm{RD}}$) is given by Eq.\,(\ref{Hl}) denotes the tunneling from the left superconductor (quantum dots) to the quantum dots (right superconductor), while $T_{d_r d_l}$ [Eq.\,(\ref{HD})] represents the tunneling from the left quantum dots (dot 1 or dot 2) to the right dots (dot 3 or dot 4)
    The arrows indicate their chronological order. 
    The five possible sequences are labeled A--E, corresponding to Eqs.\,(\ref{EJ2eA})--(\ref{EJ2eE}).
    }
    \label{figureS1}
\end{figure}

The first three conditions apply more generally to the transfer of multiple electrons through the junction, while the fourth condition ensures that the superconducting leads return to their ground-state sectors after an integer number of Cooper pairs has been transferred. 
Applying these conditions to the sixth-order process leaves five possible operator-type orderings for transferring two electrons from the left to the right superconductor, as illustrated in Fig.\,\ref{figureS1}(b) with the arrow $\rightarrow$ indicating their chronological order.
Accordingly, the five contributions to the Josephson energy can be written as
\begin{eqnarray}
E^{\rm J}_{\rm{2e},\text{A}}\left(\varphi\right) & = & \left\langle T_{\rm RD} G_{0} T_{\rm RD} G_{0} T_{d_r d_l} G_{0} T_{d_r d_l} G_{0} T_{\rm DL} G_{0} T_{\rm DL}\right\rangle +h.c.\label{EJ2eA}
\\
E^{\rm J}_{\rm{2e},\text{B}}\left(\varphi\right) & = & \left\langle T_{\rm RD} G_{0} T_{d_r d_l} G_{0} T_{\rm RD} G_{0} T_{d_r d_l} G_{0} T_{\rm DL} G_{0} T_{\rm DL} \right\rangle +h.c.\label{EJ2eB}
\\
E^{\rm J}_{\rm{2e},\text{C}}\left(\varphi\right) & = & \left\langle T_{\rm RD} G_{0} T_{d_r d_l} G_{0} T_{\rm DL} G_{0} T_{\rm RD} G_{0} T_{d_r d_l} G_{0} T_{\rm DL} \right\rangle +h.c.\label{EJ2eC}
\\
E^{\rm J}_{\rm{2e},\text{D}}\left(\varphi\right) & = & \left\langle T_{\rm RD} G_{0} T_{d_r d_l} G_{0} T_{\rm RD} G_{0} T_{\rm DL} G_{0} T_{d_r d_l} G_{0} T_{\rm DL}\right\rangle +h.c.\label{EJ2eD}
\\
E^{\rm J}_{\rm{2e},\text{E}}\left(\varphi\right) & = & \left\langle T_{\rm RD} G_{0} T_{\rm RD} G_{0} T_{d_r d_l} G_{0} T_{\rm DL} G_{0} T_{d_r d_l} G_{0} T_{\rm DL}\right\rangle +h.c.\label{EJ2eE}
\end{eqnarray}
where $\left\langle\cdots\right\rangle$ denotes the corresponding matrix element between the initial and final ground states given by Eq.\,(\ref{groundstate}). 
Although these five contributions contain the same numbers of $T_{\rm DL}$, $T_{d_r d_l}$, and $T_{\rm RD}$ operators, their different orderings generate different intermediate virtual states and, consequently, different amplitudes.
As an illustrative example, the sequence of intermediate states associated with path A [see Eq.\,(\ref{EJ2eA}) and Fig.\,\ref{figureS1}] can be represented as
\begin{equation}
|0,0\rangle \overset{T_{\rm DL}}{\longrightarrow}\left(\begin{array}{cc}
1 & 0\\
0 & 0
\end{array}\right)_{\rm L}\left\{ \begin{array}{c}
\overset{T_{\rm DL}}{\underset{\rm LAR}{\longrightarrow}}\left(\begin{array}{cc}
2 & 0\\
0 & 0
\end{array}\right)\overset{T_{d_r d_l}}{\longrightarrow}\left(\begin{array}{cc}
1 & 1\\
0 & 0
\end{array}\right)\overset{T_{d_r d_l}}{\longrightarrow}\left(\begin{array}{cc}
0 & 2\\
0 & 0
\end{array}\right)\overset{T_{\rm RD}}{\underset{\rm LAR}{\longrightarrow}}\\ 
\overset{T_{\rm DL}}{\underset{\rm CAR}{\longrightarrow}}\left(\begin{array}{cc}
1 & 0\\
1 & 0
\end{array}\right)\overset{T_{d_r d_l}}{\longrightarrow}\left(\begin{array}{cc}
0 & 1\\
1 & 0
\end{array}\right)\overset{T_{d_r d_l}}{\longrightarrow}\left(\begin{array}{cc}
0 & 1\\
0 & 1
\end{array}\right)\overset{T_{\rm RD}}{\underset{\rm CAR}{\longrightarrow}}
\end{array}\right\} \left(\begin{array}{cc}
0 & 1\\
0 & 0
\end{array}\right)_{\rm R}\overset{T_{\rm RD}}{\longrightarrow}|-2,2\rangle.
\label{diagrame}
\end{equation}
Here, the notation
\begin{equation}
\left(\begin{array}{cc}
n_{1} & n_{3}\\
n_{2} & n_{4}
\end{array}\right),
\end{equation}
specifies the electron occupations of the four quantum dots, while the subscript L/R indicates that the corresponding superconducting lead contains a single Bogoliubov quasiparticle excitation,
$\gamma^{\dagger}_{l,\bm{k},\sigma}|N_{l}\rangle$, with excitation energy $E_{l,\bm{k}}$.
Eq.\,(\ref{diagrame}) illustrates how path A evolves through its intermediate virtual states: Starting from $|0,0\rangle$, the first $T_{\rm DL}$ transfers an electron from the left superconductor to one of the two left quantum dots, producing a single quasiparticle excitation in the left superconducting lead. 
For definiteness, the electron is shown on dot 1, while the corresponding process involving dot 2 is equivalent. 
The second $T_{\rm DL}$ then transfers another electron from the left superconductor. 
The two electrons may occupy the same quantum dot, corresponding to a local Andreev reflection (LAR) process, or different quantum dots, corresponding to a crossed Andreev reflection (CAR) process. 
At the same time, the left superconducting lead returns to its ground-state sector. 
The two electrons are subsequently transferred across the junction by the two $T_{d_r d_l}$ events, thereby populating the quantum dots on the right side (dot 3 and/or dot 4). 
Finally, two $T_{\rm RD}$ events transfer the electrons into the right superconductor through the corresponding local or crossed Andreev channel, returning the entire system to the ground-state manifold with one Cooper pair transferred from left to right. Eq.\,(\ref{diagrame}) shows path A explicitly, whereas paths B--E are obtained by changing the ordering of the intermediate tunneling events shown in $\{\cdots\}$.

Following the diagram shown in Eq.\,(\ref{diagrame}) and evaluating the virtual-state contributions in the five path shown in Fig.\,\ref{figureS1}, we obtain the corresponding Josephson energy contributed by each path as \cite{rozhkov2001,PhysRevB.81.144519}
\begin{eqnarray}
E^{\rm J}_{\rm 2e,\text{A}}\left(\varphi\right) & = & -\frac{t^{2}}{2}\left[\frac{1}{2\epsilon+U_{\rm R}}\frac{1}{2\epsilon}\frac{1}{2\epsilon+U_{\rm L}}\gamma_{\rm L}\gamma_{\rm R}+ \frac{1}{\left(2\epsilon\right)^{3}} \gamma^{\prime}_{\rm L}\gamma^{\prime}_{\rm R}\right]\cos\varphi\label{EnergyJ2eA},
\\
E^{\rm J}_{\rm 2e,\text{B}}\left(\varphi\right) & \simeq & -\frac{1}{\pi\Delta}t^{2}\left[\frac{1}{2\epsilon}\frac{1}{2\epsilon+U_{L}}\gamma_{L}\gamma_{R}+\frac{1}{(2\epsilon)^{2}}\gamma^{\prime}_{L}\gamma^{\prime}_{R}\right]\cos\varphi
\label{EnergyJ2eB},
\\
E^{\rm J}_{\rm 2e,\text{C}}\left(\varphi\right) 
& \simeq & 
-\frac{t^{2}}{4\pi\Delta^{3}}\left[\gamma_{L}\gamma_{R}+\gamma^{\prime}_{L}\gamma^{\prime}_{R}f\left(\delta r_{12},\delta r_{34}\right)\right]\cos\varphi
\label{EnergyJ2eC},
\\
E^{\rm J}_{\rm 2e,\text{D}}\left(\varphi\right) & \simeq & -\frac{t^{2}}{\pi^{2}\epsilon\Delta^{2}}\left(\gamma_{L}\gamma_{R}+\gamma^{\prime}_{L}\gamma^{\prime}_{R}\right)\cos\varphi,\label{EnergyJ2eD}
\\
E^{\rm J}_{\rm 2e,\text{E}}\left(\varphi\right) & \simeq & -\frac{t^{2}}{\pi\Delta}\left(\frac{1}{2\epsilon+U_{R}}\frac{1}{2\epsilon}\gamma_{L}\gamma_{R}+\frac{1}{\left(2\epsilon\right)^{2}}\right)\cos\varphi
,\label{EnergyJ2eE}
\end{eqnarray}
where
\begin{equation}
\gamma_{l}=2\pi\rho_{l}\left\vert t_{l}\right\vert ^{2},
\label{gammaLAR}
\end{equation}
denotes the bare tunneling rate from the $l$-superconducting lead to the adjacent dots, with $\rho_{l}$ being the normal-state density of states of superconductor $l$ at the Fermi energy, while
\begin{equation}
\gamma^{\prime}_{l} = \gamma_{l}\frac{\sin\left(k_{\rm F}\delta r_{ij}\right)}{k_{\rm F}\delta r_{ij}}\exp\left(-\frac{\left\vert \delta r_{ij}\right\vert }{\pi l_{\rm sc}}\right),
\label{gammaCAR}
\end{equation}
is the rate for associated with CAR processes between the two quantum dots coupled to the same superconductor. 
Here, $\delta r_{ij}$ is the distance of the point $i$ and $j$ in the $l$ superconductor coupling to the corresponding quantum dots, $k_{\rm F}$ is the Fermi momentum, and $l_{\rm sc}$ is the superconducting coherence length. 
In Eq.\,(\ref{EnergyJ2eC}), the function
$f(\delta r_{34},\delta r_{12})$ is defined as
\begin{equation}
f(\delta r_{34},\delta r_{12})
=\frac{2}{\pi}\int^{+\infty}_{-\infty}dX\int^{+\infty}_{-\infty}dY\frac{\cos\left(\frac{\delta r_{34}}{\pi l_{sc}}Y\right)}{\left(Y^{2}+1\right)^{3/2}}\frac{\cos\left(\frac{\delta r_{12}}{\pi l_{sc}}X\right)}{\left(X^{2}+1\right)^{3/2}}\frac{\exp\left(\frac{\delta r_{12}}{\pi l_{sc}}\right)\exp\left(\frac{\delta r_{34}}{\pi l_{sc}}\right)}{\left(Y^{2}+1\right)^{1/2}+\left(X^{2}+1\right)^{1/2}},
\end{equation}
where $Y=E_q/\Delta$ and $X=E_k/\Delta$ are dimensionless quasiparticle energies. The function $f(\delta r_{34},\delta r_{12})$ exhibits oscillatory decay as a function of the interdot distances $\delta r_{34}$ and $\delta r_{12}$.

In large-superconducting-gap limit with $\Delta\rightarrow\infty$, Eqs.~(\ref{EnergyJ2eB})--(\ref{EnergyJ2eE}) vanish, and the single-Cooper-pair contribution to the Josephson energy is determined solely by path A,
\begin{equation}
E^{\rm J}_{\rm 2e}(\varphi)
=
\lim_{\Delta\rightarrow\infty}E^{\rm J}_{\rm 2e,\text{A}}(\varphi) 
=
-\frac{t^{2}}{2}\left[\frac{1}{2\epsilon+U_{\rm R}}\frac{1}{2\epsilon}\frac{1}{2\epsilon+U_{\rm L}}\gamma_{\rm L}\gamma_{\rm R}+ \frac{1}{\left(2\epsilon\right)^{3}} \gamma^{\prime}_{\rm L}\gamma^{\prime}_{\rm R}\right]\cos\varphi.
\label{EJ2eLargeDelta}
\end{equation}
where the first and second terms in the square brackets originate from local and crossed Andreev reflection processes, respectively, and are characterized by the local and nonlocal tunneling rates $\gamma_{l}$ and $\gamma^{\prime}_{l}$.
For the configuration considered in Fig. 1 in the main text, we take $\gamma^{\prime}_{\rm R}=0$, such that Eq.~(\ref{EnergyJ2eA}) reduces to
\begin{equation}
E^{\rm J}_{\rm 2e,\text{A}}\left(\varphi\right)=-\frac{t^{2}}{2}\frac{1}{2\epsilon+U_{\rm R}}\frac{1}{2\epsilon}\frac{1}{2\epsilon+U_{\rm L}}\gamma_{\rm L}\gamma_{\rm R}\cos\varphi.
\label{EJ2eAnoCAR}
\end{equation}
The remaining single-Cooper-pair contribution therefore relies entirely on local Andreev processes. It is consequently suppressed by the large charging energy $U_{L}$ and vanishes in the limit $U_{L}\rightarrow\infty$,
\begin{equation}
\lim_{\Delta,U_{\rm L}\rightarrow\infty,\gamma^{\prime}_{\rm R}=0}E^{\rm J}_{\rm 2e}\left(\varphi\right)=0.
\label{EJ2eLimit}
\end{equation}
Hence, in the combined limit of an infinite superconducting gap, strong Coulomb repulsion  on the left pair of quantum dots, and vanishing crossed Andreev coupling on the right, the conventional single-Cooper-pair contribution to the Josephson energy is completely suppressed.

\subsection{Double Cooper-pair contribution}

The suppression of the single-Cooper-pair contribution in Eq.\,(\ref{EJ2eLimit}) motivates us to examine higher-order tunneling processes. We therefore consider the coherent transfer of two Cooper pairs, corresponding to four electrons, between the superconducting leads. The resulting contribution to the Josephson energy is
\begin{equation}
E^{\rm J}_{\rm 4e}\left(\varphi\right)
=
2\mathrm{Re}\left[
\langle-4,4|
\mathcal{T}^{\left(12\right)}
|0,0\rangle
\right],
\end{equation}
where the operator $\mathcal{T}^{\left(12\right)}$ represents the twelfth-order contribution in the tunneling Hamiltonian,
\begin{equation}
\mathcal{T}^{\left(12\right)}
=
T\left(G_{0}T\right)^{11}.
\end{equation}
The transfer of four electrons requires four tunneling events generated by each of $T_{\rm DL}$, $T_{d_r d_l}$, and $T_{\rm RD}$, and involves eleven intermediate virtual states.
Formally, twelve individually labeled tunneling events generate $12!$ possible permutations, which is then reduced to 468 allowed operator sequences after applying Rules 1--4.
Furthermore, in the large-superconducting-gap limit, many of these sequences are suppressed because their intermediate-state energy denominators contain Bogoliubov quasiparticle excitation energies over nonadjacent tunneling events, yielding contributions that vanish as inverse powers of $\Delta$, which is similar to Eq.\,(\ref{EnergyJ2eB}-\ref{EnergyJ2eE}). 
The leading contribution relevant to the regime considered here is generated by the ordering
\begin{equation}
\left\langle
T_{\rm RD}\left(G_{0}T_{\rm RD}\right)^{3}
\left(G_{0}T_{d_r d_l}\right)^{2}
\left(G_{0}T_{\rm DL}\right)^{2}
\left(G_{0}T_{d_r d_l}\right)^{2}
\left(G_{0}T_{\rm DL}\right)^{2}
\right\rangle
+h.c..
\label{E4eleading}
\end{equation}
Depending on whether the two Cooper-pair transfers at each superconducting interface proceed through local or crossed Andreev processes, the Josephson energy correspouding to Eq.\,(\ref{E4eleading}) can be decomposed as
\begin{equation}
E^{\rm J}_{\rm 4e}\left(\varphi\right)
\simeq
E^{\text{R-LAR}}_{\text{L-LAR}}
+
E^{\text{R-LAR}}_{\text{L-CAR}}
+
E^{\text{R-CAR}}_{\text{L-LAR}}
+
E^{\text{R-CAR}}_{\text{L-CAR}}.
\label{EJ4esimple}
\end{equation}
Here, $E^{\rm R-\beta}_{\rm L-\alpha}$, with $\alpha,\beta\in\{\mathrm{LAR},\mathrm{CAR}\}$, denotes the contribution in which the transfer between the left superconductor and dots 1 and 2 proceeds through process $\alpha$, while that between dots 3 and 4 and the right superconductor proceeds through process $\beta$.
As an example, we consider $E^{\text{R-LAR}}_{\text{L-CAR}}$, for which two crossed Andreev processes occur at the left interface and two local Andreev processes occur at the right interface. 
The corresponding sequence of virtual states can be represented as
\begin{eqnarray}
|0,0\rangle
&\overset{T_{\rm DL}}{\longrightarrow}&
\left(\begin{array}{cc}
1 & 0\\
0 & 0
\end{array}\right)_{L}
\overset{T_{\rm DL}}{\underset{\rm CAR}{\longrightarrow}}
\left(\begin{array}{cc}
1 & 0\\
1 & 0
\end{array}\right)
\overset{T_{d_r d_l}}{\longrightarrow}
\left(\begin{array}{cc}
0 & 1\\
1 & 0
\end{array}\right)
\overset{T_{d_r d_l}}{\longrightarrow}
\left(\begin{array}{cc}
0 & 1\\
0 & 1
\end{array}\right)
\nonumber\\
&\overset{T_{\rm DL}}{\longrightarrow}&
\left(\begin{array}{cc}
1 & 1\\
0 & 1
\end{array}\right)_{\rm L}
\overset{T_{\rm DL}}{\underset{CAR}{\longrightarrow}}
\left(\begin{array}{cc}
1 & 1\\
1 & 1
\end{array}\right)
\overset{T_{d_r d_l}}{\longrightarrow}
\left(\begin{array}{cc}
0 & 2\\
1 & 1
\end{array}\right)
\overset{T_{d_r d_l}}{\longrightarrow}
\left(\begin{array}{cc}
0 & 2\\
0 & 2
\end{array}\right)
\label{diagrameE4e}\\
&\overset{T_{\rm RD}}{\longrightarrow}&
\left(\begin{array}{cc}
0 & 1\\
0 & 2
\end{array}\right)_{\rm R}
\overset{T_{\rm RD}}{\underset{\rm LAR}{\longrightarrow}}
\left(\begin{array}{cc}
0 & 0\\
0 & 2
\end{array}\right)
\overset{T_{\rm RD}}{\longrightarrow}
\left(\begin{array}{cc}
0 & 0\\
0 & 1
\end{array}\right)_{\rm R}
\overset{T_{\rm RD}}{\underset{\rm LAR}{\longrightarrow}}
|-4,4\rangle.
\nonumber
\end{eqnarray}
Starting from $|0,0\rangle$, the first $T_{\rm DL}$ transfers an electron from the left superconductor to one of the two left quantum dots, producing a single Bogoliubov quasiparticle excitation in the left superconducting lead. 
The second $T_{\rm DL}$ completes a CAR process by transferring the second electron to the other left dot and simultaneously returning the left superconductor to its ground-state sector. 
The two electrons are subsequently transferred through the two $T_{d_r d_l}$ processes to dots 3 and 4, respectively, as shown in the first line of Eq.~(\ref{diagrameE4e}).
The same sequence $T_{d_r d_l}T_{d_r d_l}T_{\rm DL}T_{\rm DL}$ is then repeated, resulting in double occupation of both right quantum dots, as shown in the second line of Eq.~(\ref{diagrameE4e}). Importantly, once the first pair has been transferred through CAR, the second pair must also enter the left dots through CAR for this particular sequence.
A subsequent LAR process would, otherwise, doubly occupy one of the left dots, so that the corresponding $T_{d_r d_l}$ processes would attempt to transfer additional electrons into an already occupied single-orbital state on the corresponding right dot, yielding a vanishing matrix element. 
Finally, four $T_{\rm RD}$ operations remove the four electrons from the right dots through two successive LAR processes and transfer two Cooper pairs into the right superconductor.
The Josephson-energy contribution associated with this process is
\begin{equation}
E^{\text{R-LAR}}_{\text{L-CAR}}
=
\frac{t^{4}}{32}\frac{1}{(2\epsilon+U_{R})^{2}}\frac{1}{4\epsilon+U_{R}}\frac{1}{(2\epsilon)^{4}}\gamma^{2}_{R}\gamma^{\prime2}_{L}\cos2\varphi.
\end{equation}
Using the same procedure, the remaining three contributions in Eq.\,(\ref{EJ4esimple}) are
\begin{eqnarray}
E^{\text{R-LAR}}_{\text{L-LAR}}
&=&
\frac{t^{4}}{16}\frac{1}{(2\epsilon+U_{R})^{3}}\frac{1}{4\epsilon+U_{R}}\frac{1}{4\epsilon+U_{R}+U_{L}}\frac{1}{2\epsilon}\frac{1}{2\epsilon+U_{L}}\gamma^{2}_{L}\gamma^{2}_{R}\cos2\varphi,
\\
E^{\text{R-CAR}}_{\text{L-LAR}}
&=&
\frac{t^{4}}{8}\frac{1}{\left(2\epsilon\right)^{2}}\frac{1}{2(2\epsilon+U_{R})^{2}}\frac{1}{4\epsilon+U_{R}}\frac{1}{4\epsilon+U_{R}+U_{L}}\frac{1}{2\epsilon+U_{L}}\gamma^{2}_{L}\gamma^{\prime2}_{R}\cos2\varphi,
\\
E^{\text{R-CAR}}_{\text{L-CAR}}
&=&
\frac{t^{4}}{8}\times\frac{1}{4\left(2\epsilon\right)^{5}}\frac{1}{(2\epsilon+U_{R})}\frac{1}{4\epsilon+U_{R}}\gamma^{\prime2}_{L}\gamma^{\prime2}_{R}\cos2\varphi.
\end{eqnarray}
We now consider the parameter regime relevant to our setup. 
In the absence of CAR between the right superconductor and dots 3 and 4 ($\gamma^{\prime}_{\rm R}=0$), the contributions involving right-side CAR vanish:
\begin{equation}
E^{\text{R-CAR}}_{\text{L-LAR}}
=
E^{\text{R-CAR}}_{\text{L-CAR}}
=0.
\end{equation}
Moreover, in the strong-Coulomb-blockade limit on the left dots ($U_{\rm L}\rightarrow\infty$) the contribution involving left-side LAR is suppressed,
\begin{equation}
E^{\text{R-LAR}}_{\text{L-LAR}}\rightarrow0.
\end{equation}
Consequently, the leading two-Cooper-pair contribution is
\begin{equation}
E^{\rm J}_{\rm 4e}\left(\varphi\right)
\simeq
E^{\text{R-LAR}}_{\text{L-CAR}},
\end{equation}
which remains finite in the combined limit of large $\Delta$ and $U_{\rm L}$ and vanishing right-side CAR. 
Physically, this contribution transfers two Cooper pairs through a process involving CAR at the left interface and LAR at the right interface.

More generally, the Josephson energy can be organized according to the number of Cooper pairs transferred coherently,
\begin{equation}
E^{\rm J}\left(\varphi\right)
=
E^{\rm J}_{\rm 2e}\left(\varphi\right)
+
E^{\rm J}_{\rm 4e}\left(\varphi\right)
+
E^{\rm J}_{\rm 6e}\left(\varphi\right)
+\cdots.
\end{equation}
Retaining the two lowest contributions gives 
\begin{eqnarray}
E^{\rm J}\left(\varphi\right)
\simeq
E^{\rm J}_{\rm 2e}\left(\varphi\right)
+
E^{\rm J}_{\rm 4e}\left(\varphi\right)
=
-\Gamma_{\rm 2e}\cos\varphi
+
\Gamma_{\rm 4e}\cos2\varphi,
\end{eqnarray}
where
\begin{eqnarray}
\Gamma_{\rm 2e}
&=&
\frac{t^{2}\gamma_{L}\gamma_{R}}{4\epsilon\left(2\epsilon+U_{R}\right)\left(2\epsilon+U_{L}\right)},
\label{Gamma2e}
\\
\Gamma_{\rm 4e}
&=&
\frac{t^{4}\gamma^{\prime2}_{L}\gamma^{2}_{R}}{32(2\epsilon)^{4}(2\epsilon+U_{R})^{2}\left(4\epsilon+U_{R}\right)}
\label{Gamma4e}
.
\end{eqnarray}
The corresponding Josephson current is
\begin{equation}
I^{\rm J}
=
\frac{2e}{\hbar}
\partial_{\varphi}E^{\rm J}\left(\varphi\right)
\simeq
I_{\rm 2e}+I_{\rm 4e},
\end{equation}
with
\begin{eqnarray}
I_{\rm 2e}
&=&
\frac{2e}{\hbar}
\Gamma_{\rm 2e}
\sin\varphi,
\\
I_{\rm 4e}
&=&
-\frac{4e}{\hbar}
\Gamma_{\rm 4e}
\sin2\varphi.
\end{eqnarray}
Therefore, in the limit $U_{\rm L}\rightarrow\infty$, the conventional single-Cooper-pair contribution is suppressed as $\Gamma_{\rm 2e}\rightarrow0$, whereas the two-Cooper-pair contribution remains finite. We consequently obtain
\begin{equation}
\lim_{U_{\rm L}\rightarrow\infty}I^{\rm J}
\simeq
I_{\rm 4e}
=
\frac{4e}{\hbar}
\Gamma_{\rm 4e}
\sin2\varphi.
\label{IJULlmit}
\end{equation}
The resulting $\pi$-periodic current-phase relation therefore signals a Josephson current dominated by the coherent transfer of two Cooper pairs.

\section{Josephson diode effect}

\subsection{Effect of magnetic flux and junction asymmetries}

The magnetic flux penetrating the four-quantum-dot system is incorporated into the interdot hopping amplitudes through the Peierls substitution,
\begin{eqnarray}
t_{13} \rightarrow  t_{13}e^{i\theta_{13}},
\quad
t_{24} \rightarrow  t_{24}e^{i\theta_{24}},
\end{eqnarray}
where
\begin{equation}
\theta_{ij}=\frac{e}{\hbar}\int_j^i
\bm{A}\cdot d\bm{l}
\end{equation}
is the Peierls phase accumulated during electron tunneling from dot $j$ to dot $i$, with $\bm{A}$ denoting the magnetic vector potential.
The dimensionless flux phase accumulated in the
four quantum dots is 
\begin{eqnarray}
\tilde{\Phi}
= \frac{e}{\hbar}\Phi
=\frac{e}{\hbar}\oint\bm{A}\cdot d\bm{l} 
= \frac{e}{\hbar}
\left[
-A_y l_{12}+A_x l_{24}
+A_y l_{34}-A_x l_{13}
\right],
\end{eqnarray}
where $\Phi$ is the  magnetic flux threading the closed loop formed by dots and the vector-potential components in the last expression are understood to be evaluated along their respective tunneling paths.
Choosing the Landau gauge with $A_y=0$, the phase difference between the two interdot tunneling paths becomes
\begin{equation}
\theta_{24}-\theta_{13}
=\tilde{\Phi}
=2\pi\frac{\Phi}{\Phi_e},
\end{equation}
where $\Phi_e=h/e$ is the single-electron flux quantum. For a symmetric choice of the two Peierls phases, we set
\begin{eqnarray}
\theta_{13} = -\frac{\tilde{\Phi}}{2},
\quad
\theta_{24} = \frac{\tilde{\Phi}}{2}.
\end{eqnarray}
Then following Eq.\,(\ref{EnergyJ2eA}), the leading single-Cooper-pair contribution to the Josephson energy becomes
\begin{eqnarray}
E^{\rm J}_{\rm 2e,\text{A}}
&=&-\frac{
t_{13}^2e^{-i\tilde{\Phi}}
+t_{24}^2e^{i\tilde{\Phi}}
}{4}
\frac{\gamma_{\rm L}\gamma_{\rm R}}
{2\epsilon(2\epsilon+U_{\rm L})(2\epsilon+U_{\rm R})}
e^{i\varphi}
+\mathrm{H.c.}
\nonumber\\
&=&-\frac{t_{13}^2}{2}
\frac{\gamma_{\rm L}\gamma_{\rm R}}
{2\epsilon(2\epsilon+U_{\rm L})(2\epsilon+U_{\rm R})}
\left[
\cos(\varphi-\tilde{\Phi})
+\frac{t_{24}^2}{t_{13}^2}
\cos(\varphi+\tilde{\Phi})
\right].
\label{E2eAflux}
\end{eqnarray}
Introducing the hopping asymmetry parameter
\begin{equation}
\mathcal{A}\equiv\frac{t_{24}^2}{t_{13}^2},
\end{equation}
with $t_{13}^2\equiv|t|^2$, and using
Eq.\,(\ref{Gamma2e}), we can express
Eq.\,(\ref{E2eAflux}) in the compact form
\begin{equation}
E^{\rm J}_{\rm{2e,A}}
=-\Gamma_{\rm 2e}P_{\Phi}
\cos(\varphi-\varphi_{\Phi}),
\end{equation}
where
\begin{eqnarray}
P_{\Phi}
&=&
\sqrt{
[(1+\mathcal{A})\cos\tilde{\Phi}]^2+
[(1-\mathcal{A})\sin\tilde{\Phi}]^2
},\\
\varphi_{\Phi}
&=&
\arctan \left(\frac{1-\mathcal{A}}{1+\mathcal{A}} \tan\tilde{\Phi} \right)
.
\end{eqnarray}
Here, $P_{\Phi}$ characterizes the flux-dependent modulation
of the Josephson coupling, while $\varphi_{\Phi}$ denotes the
corresponding anomalous phase shift.

The resulting Josephson current is
\begin{eqnarray}
I_{\rm 2e} = \frac{2e}{\hbar}
\frac{\partial E^{\rm J}_{\rm{2e,A}}}{\partial\varphi} = \frac{2e}{\hbar}
\Gamma_{\rm 2e}P_{\Phi}
\sin(\varphi-\varphi_{\Phi}).
\end{eqnarray}
Consequently, the interplay between magnetic flux and
asymmetric interdot hopping produces an anomalous
phase shift, giving rise to a $\varphi_0$-junction with
an equilibrium phase $\varphi_{\Phi}$ when only the
single-Cooper-pair contribution is considered.

We next examine the influence of magnetic flux on the double-Cooper-pair contribution. The corresponding Josephson energy is written as
\begin{eqnarray}
E^{\rm J}_{\rm 4e}
 & = & \mathcal{A}\Gamma_{\rm 4e}\cos2\varphi,
\end{eqnarray}
which gives the Josephson current
\begin{eqnarray}
I_{\rm 4e} & = & -\frac{4e}{\hbar}\mathcal{A}\Gamma_{\rm 4e}\sin2\varphi.
\end{eqnarray}
Combining the single- and double-Cooper-pair contributions, the total Josephson current becomes
\begin{eqnarray}
I^{\rm J}  \simeq  I_{\rm 2e}+I_{\rm 4e} = \frac{2e}{\hbar}\Gamma_{\rm 2e}P_{\Phi}\sin(\varphi-\varphi_{\Phi})-\frac{4e}{\hbar}\mathcal{A}\Gamma_{\rm 4e}\sin2\varphi.
\end{eqnarray}
which is a superposition of a flux-controlled $\varphi_{0}$-junction and a junction dominated by the second Josephson harmonic. 
Importantly, the single-Cooper-pair contribution alone remains reciprocal: magnetic flux and hopping asymmetry modify its amplitude and shift its equilibrium phase, but do not generate a Josephson diode effect. 
By contrast, the double-Cooper-pair contribution introduces a higher harmonic into the current-phase relation. 
Its interference with the phase-shifted single-Cooper-pair current can break the symmetry between the positive and negative critical currents, thereby giving rise to a Josephson diode effect. 

 \subsection{Maximized diode efficiency condition}

The Josephson diode effect originates from the interference between the single- and double-Cooper-pair currents. In the regime where the first harmonic dominates the current-phase relation, the diode asymmetry is enhanced when the anomalous phase shift satisfies
\begin{equation}
\varphi^c_{\Phi}=\pi/4+n\pi/2,
\end{equation}
where $n\in\mathbb{Z}$. 
Using the expression for $\varphi_{\Phi}$ derived above, the corresponding critical hopping asymmetry $\mathcal{A}_{c}$ and magnetic flux $\tilde{\Phi}_{c}$ satisfy
\begin{equation}
\varphi_{\Phi}=\arctan\left(\frac{1-\mathcal{A}_{c}}{1+\mathcal{A}_{c}}\tan\tilde{\Phi}_{c}\right)
\Rightarrow
\frac{1-\mathcal{A}_{c}}{1+\mathcal{A}_{c}}\tan\tilde{\Phi}_{c}=\pm1.
\end{equation}
Equivalently,
\begin{equation}
(1-\mathcal{A}_{c})\sin\tilde{\Phi}_{c}
=
\pm(1+\mathcal{A}_{c})\cos\tilde{\Phi}_{c}.
\end{equation}
Under this condition, the flux-dependent modulation factor becomes
\begin{equation}
P^c_{\Phi}=\sqrt{2}(1+\mathcal{A}_c)|\cos\tilde{\Phi}_{c}|.
\end{equation}
For the branch in which the single- and double-Cooper-pair currents interfere constructively at $\varphi=3\pi/4$ and destructively at $\varphi=-\pi/4$, the corresponding positive and negative critical currents are
\begin{eqnarray}
I^{\rm J}_{+c} & = & \frac{2e}{\hbar}\Gamma_{2e}P_{\Phi}+\frac{4e}{\hbar}\mathcal{A}\Gamma_{4e},\\
|I^{\rm J}_{-c}| & = & \frac{2e}{\hbar}\Gamma_{2e}P_{\Phi}-\frac{4e}{\hbar}\mathcal{A}\Gamma_{4e}.
\end{eqnarray}
Consequently, the diode efficiency takes the form
\begin{eqnarray}
\eta_{\rm max}
&=&
\max\left|
\frac{I^{\rm J}_{+c}-|I^{\rm J}_{-c}|}
{I^{\rm J}_{+c}+|I^{\rm J}_{-c}|}
\right|
=
\frac{\mathcal{A}_{c}}{P_{\Phi_{c}}}
\frac{2\Gamma_{\rm 4e}}{\Gamma_{\rm 2e}}.
\end{eqnarray}
Thus, the double-Cooper-pair contribution is characterized by a Josephson diode: 
without the contribution from the double Cooper pair from the Cooper pair splitting ($\Gamma_{\rm 4e} =0$), the maximum diode efficiency is zero $\eta_{\rm max} =0 $.

We also note the at small $\mathcal{A}$ and high $U_{\rm L}$, $P^c_{\Phi} \simeq 1$, the suppression of the Cooper-pair current is due to the high $U_{\rm L}$ (small $\Gamma_{\rm 2e}$).

\section{The $\Delta\rightarrow\infty$ limit}

For subgap transport, we consider the infinite-superconducting-gap
limit, $\Delta\rightarrow\infty$, in which quasiparticle excitations
in the superconducting leads become inaccessible and the lead
degrees of freedom can be integrated out.
The resulting effective Hamiltonian is given by
\begin{equation}
H^{\prime}_{\rm JJ}
=H_{\rm d}+H_{\rm Sd}+T_{\rm D},
\label{Htotal}
\end{equation}
where $H_{\rm d}$ describes the four-quantum-dot system [Eq.~(\ref{Hd})] and $T_{\rm D}$ accounts for interdot tunneling [Eq.~(\ref{HD})].
$H_{\rm Sd}$ represents the proximity-induced superconducting coupling between the quantum dots and the superconducting leads, which can be expressed in terms of the creation and
annihilation of Cooper pairs in the quantum dots through local and crossed Andreev processes:
\begin{eqnarray}
H_{\rm Sd}
&=&
\sum_i \frac{\gamma_i}{2} d_{i\uparrow}^{\dagger}d_{i\downarrow}^{\dagger}
+
\frac{\gamma_{12}}{2}
\left(
d_{1\uparrow}^{\dagger}d_{2\downarrow}^{\dagger}
-d_{1\downarrow}^{\dagger}d_{2\uparrow}^{\dagger}
\right)
+\frac{\gamma_{34}}{2}
\left(
d_{3\uparrow}^{\dagger}d_{4\downarrow}^{\dagger}
-d_{3\downarrow}^{\dagger}d_{4\uparrow}^{\dagger}
\right)
+\mathrm{H.c.},
\label{HSd}
\end{eqnarray}
where
\begin{eqnarray}
\gamma_1=\gamma_2=\gamma_{\rm L}e^{i\varphi_{\rm L}},
\quad
\gamma_3=\gamma_4=\gamma_{\rm R}e^{i\varphi_{\rm R}}
\end{eqnarray}
are the local pairing amplitudes associated with local Andreev reflection between each superconducting lead and its adjacent quantum dots. 
The nonlocal pairing amplitudes,
\begin{eqnarray}
\gamma_{12}=\gamma_{\rm L}^{\prime}e^{i\varphi_{\rm L}},
\quad
\gamma_{34}=\gamma_{\rm R}^{\prime}e^{i\varphi_{\rm R}},
\end{eqnarray}
describe crossed Andreev reflection (CAR), in which a Cooper pair is split between two quantum dots coupled to the same superconducting lead.

Although the pairing Hamiltonian $H_{\rm Sd}$ does not conserve particle number, it changes the electron number only by two, thereby preserving fermion parity. 
Since $H_{\rm d}$ and $T_{\rm D}$ also conserve fermion parity, the effective Hamiltonian in Eq.~(\ref{Htotal}) can be decomposed into independent even- and odd-parity sectors. 
The full Hilbert space of the four-quantum-dot system has dimension $4^4=256$, comprising $128$ even-parity and $128$ odd-parity states. 
In the parameter regime considered here, the ground state belongs to the even-parity sector so that We therefore restrict our analysis to this sector, reducing the Hamiltonian dimension from $256$ to $128$.

To calculate the Josephson current, we choose the superconducting phases as $\varphi_{\rm L}=-\varphi/2$ and $\varphi_{\rm R}=\varphi/2$, where $\varphi=\varphi_{\rm R}-\varphi_{\rm L}$ denotes the superconducting phase difference across the junction. 
The Josephson-current operator is obtained by differentiating the effective Hamiltonian with respect to the phase difference,
\begin{equation}
\hat{I}
=
\frac{2e}{\hbar}\partial_{\varphi}H^{\prime}_{\rm JJ}
=
\frac{2e}{\hbar}\partial_{\varphi}H_{\rm Sd}.
\end{equation}
At zero temperature, the equilibrium Josephson current is evaluated from the ground-state expectation value,
\begin{equation}
I(\varphi)
=
\langle\psi_{\rm gs}|\hat{I}|\psi_{\rm gs}\rangle,
\end{equation}
where $|\psi_{\rm gs}\rangle$ is the ground state obtained
by diagonalizing the effective Hamiltonian in
Eq.~(\ref{Htotal}). 
Finally, the positive and negative critical current
are defined as
\begin{equation}
I_{+c}=\max_{\varphi} I(\varphi)
\quad
I_{-c}=|\min_{\varphi}I(\varphi)|
,
\end{equation}
respectively.


%